\documentclass[sigconf]{acmart}

\AtBeginDocument{%
  \providecommand\BibTeX{{%
    \normalfont B\kern-0.5em{\scshape i\kern-0.25em b}\kern-0.8em\TeX}}}

\copyrightyear{2021} 
\acmYear{2021} 
\setcopyright{acmcopyright}\acmConference[KDD '21]{Proceedings of the 27th ACM SIGKDD Conference on Knowledge Discovery and Data Mining}{August 14--18, 2021}{Virtual Event, Singapore}
\acmBooktitle{Proceedings of the 27th ACM SIGKDD Conference on Knowledge Discovery and Data Mining (KDD '21), August 14--18, 2021, Virtual Event, Singapore}
\acmPrice{15.00}
\acmDOI{10.1145/3447548.3467434}
\acmISBN{978-1-4503-8332-5/21/08}

\usepackage{booktabs}
\usepackage{soul}
\usepackage{url}
\usepackage{hyperref}
\usepackage[utf8]{inputenc}
\usepackage{caption}
\usepackage{graphicx}
\usepackage{amsmath}
\usepackage{booktabs}
\usepackage{algorithm}
\usepackage{algorithmic}

\usepackage{amssymb}
\usepackage{amsmath,amsfonts}

\usepackage{algorithmic}
\usepackage{graphicx}
\usepackage{textcomp}
\usepackage{xcolor}
\def\BibTeX{{\rm B\kern-.05em{\sc i\kern-.025em b}\kern-.08em
    T\kern-.1667em\lower.7ex\hbox{E}\kern-.125emX}}

\usepackage{tikz}
\usepackage[caption=false]{subfig}
\usepackage[normalem]{ulem}
\usepackage{algorithm}
\usepackage{array}
\usepackage{color}
\usepackage{amsmath,amssymb}
\usepackage{multicol}
\usepackage{multirow}
\usepackage{array}
\usepackage{bm}
\usepackage{mathtools}
\usepackage{amssymb}
\usepackage{url}
\usepackage{appendix}

\newcolumntype{I}{!{\vrule width 2pt}}
\newlength\savedwidth
\newcommand\whline{\noalign{\global\savedwidth\arrayrulewidth
\global\arrayrulewidth 2pt}%
\hline
\noalign{\global\arrayrulewidth\savedwidth}}
\begin{document}
\fancyhead{}

\title{TUTA: Tree-based Transformers for Generally Structured Table Pre-training}

\author{Zhiruo Wang*}
\affiliation{
  \institution{Carnegie Mellon University}
  \streetaddress{}
  \city{}
  \country{}}
\email{zhiruow@andrew.cmu.edu}
\thanks{* Equal contribution. Work done during Zhiruo's internship at Microsoft Research.}

\author{Haoyu Dong* \texorpdfstring{$\dagger$}{Lg}}
\affiliation{
  \institution{Microsoft Research}
  \streetaddress{}
  \city{}
  \country{}}
\email{hadong@microsoft.com}
\thanks{\texorpdfstring{$\dagger$}{Lg} Corresponding author.}

\author{Ran Jia}
\affiliation{
  \institution{Microsoft Research}
  \streetaddress{}
  \city{}
  \country{}}
\email{jia.ran@microsoft.com}

\author{Jia Li}
\affiliation{
  \institution{Peking university}
  \streetaddress{}
  \city{}
  \country{}}
\email{lijiaa@pku.edu.cn}

\author{Zhiyi Fu}
\affiliation{
  \institution{Peking University}
  \streetaddress{}
  \city{}
  \country{}}
\email{ypfzy@pku.edu.cn}

\author{Shi Han}
\affiliation{
  \institution{Microsoft Research}
  \streetaddress{}
  \city{}
  \country{}}
\email{shihan@microsoft.com}

\author{Dongmei Zhang}
\affiliation{
  \institution{Microsoft Research}
  \streetaddress{}
  \city{}
  \country{}}
\email{dongmeiz@microsoft.com}

\begin{abstract}
Tables are widely used with various structures to organize and present data. 
Recent attempts on table understanding mainly focus on relational tables, yet overlook to other common table structures.
In this paper, we propose TUTA, a unified pre-training architecture for understanding generally structured tables.
Noticing that understanding a table requires spatial, hierarchical, and semantic information, we enhance transformers with three novel structure-aware mechanisms.
First, we devise a unified tree-based structure, called a bi-dimensional coordinate tree, to describe both the spatial and hierarchical information of generally structured tables.
Upon this, we propose tree-based attention and position embedding to better capture the spatial and hierarchical information.
Moreover, we devise three progressive pre-training objectives to enable representations at the token, cell, and table levels.
We pre-train TUTA on a wide range of unlabeled web and spreadsheet tables and fine-tune it on two critical tasks in the field of table structure understanding: cell type classification and table type classification.
Experiments show that TUTA is highly effective, achieving state-of-the-art on five widely-studied datasets.
\end{abstract}



\maketitle

\section{Introduction} \label{sec:intro}

\begin{figure}[t]
    \begin{center}
    \includegraphics[width=3.35in]{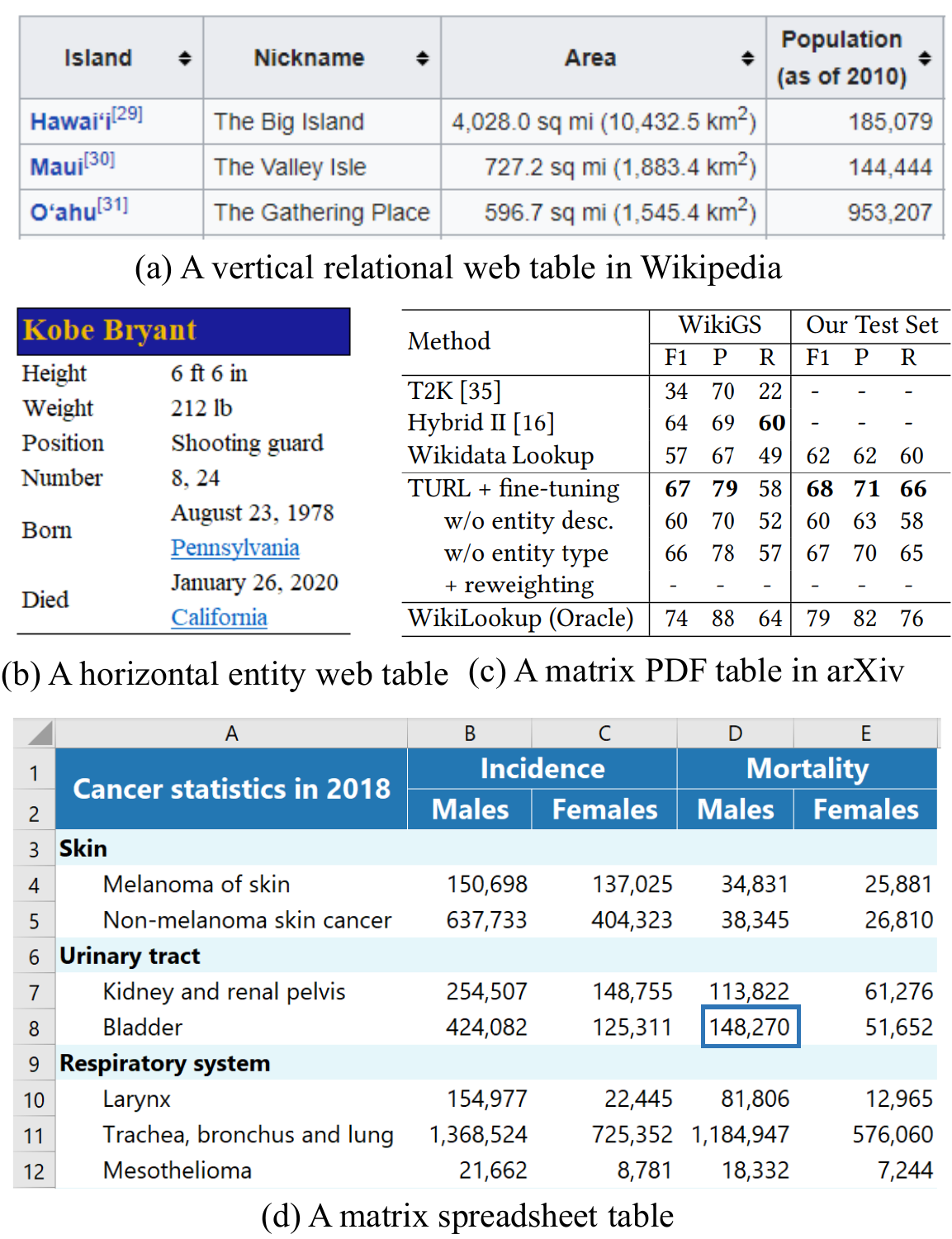}
    \end{center}
\vspace{-0.3cm}
\caption{Examples of generally structured tables. (a) is a relational web table. (b) is an entity web table. (c)(d) are matrix PDF and spreadsheet tables.}
\label{fig:realtable}
\vspace{-0.3cm}
\end{figure}

Table, as a key structure to organize and present data, is widely used in webpages, spreadsheets, PDFs, and more.
Unlike sequential NL text, tables usually arrange cells in bi-dimensional matrices with stylistic formatting such as border, merge, alignment, and bold font. 
As Figure~\ref{fig:realtable} shows, \textbf{tables are flexible with various structures}, such as relational, entity and matrix tables~\cite{nishida2017understanding}. To present data more effectively in the bi-dimensional space, real-world tables often use hierarchical structures~\cite{chen2014integrating,chen2013automatic,dou2018expandable}. For example, Figure~\ref{fig:realtable}~(d) shows a matrix spreadsheet table with both hierarchical top and left headers (``B1:E'' and ``A3:A2''), shaped by merged cells and indents respectively, to organize data in a compact and hierarchical way for easy look-up or side-by-side comparison. 

Tables store a great amount of high-value data and gain increasing attention from research community. There is a flurry of research on table understanding tasks, including entity linking in tables~\cite{bhagavatula2015tabel,ritze2017matching,zhao2019auto,deng2020turl}, column type identification of tables~\cite{chen2019colnet,guo2020web}, answering natural language questions over tables~\cite{pasupat2015compositional,yu2018spider,herzig2020tapas,yin2020tabert}, and generating data analysis for tables~\cite{zhou2020table2analysis}.
\textbf{However, the vast majority of these work only focuses on relational tables}, which only account for $0.9\%$ of commonly crawled web tables\footnote{http://webdatacommons.org/webtables/} and $22.0\%$ of spreadsheet tables~\cite{chen2014integrating}. They overlook other widely used table types such as matrix tables and entity tables. This leads to a large gap between cutting-edge table understanding techniques and variously structured real-world tables. Therefore, it is important to enable table understanding for variously structured tables and make a critical step to mitigate this gap. There are several attempts~\cite{dong2019semantic,gol2019tabular,koci2019deco,chen2014integrating,chen2013automatic} on identifying table hierarchies and cell types to extract relational data from variously structured tables. However, \textbf{labeling such structural information is very time-consuming and labor-intensive}, hence greatly challenges machine learning methods that are rather data-hungry.  

Motivated by the success of large-scale pre-trained language models (LMs)~\cite{devlin2018bert,radford2018improving} in numerous NL tasks, one promising way to mitigate the label-shortage challenge is self-supervised pre-training on large volumes of unlabeled tables.~\cite{herzig2020tapas,yin2020tabert} target question answering over relational tables via joint pre-training of tables and their textual descriptions.~\cite{deng2020turl} attempts to pre-train embeddings on relational tables to enhance table knowledge matching and table augmentation. However, \textbf{these pre-training methods still only target relational web tables} due to their simple structures. In relational tables, each column is homogeneous and described by a column name, so~\cite{yin2020tabert} augments each data cell with its corresponding column name, and~\cite{deng2020turl} enforces each cell aggregate information from only its row and column. 
But clearly, these methods are not suitable for tables of other structures. 
For example, Figure~\ref{fig:realtable}~(d) shows a hierarchical matrix table, where cell ``D'' (`148,270') is jointly described by cells in multiple rows and columns: ``D'' (`Mortality'), ``D2'' (`Males'), ``A'' (`Urinary tract') and ``A8'' (`Bladder'), constituting to a relational tuple (`Mortality', `Males', `Urinary tract', `Bladder', `148,270')~\cite{chen2014integrating}. 
Such structural layouts can greatly help human readers to understand tables. Simply treating tables as relational will lose valuable structural information.

Hence, \textbf{we aim to propose a structure-aware method for generally structured table pre-training.}
Fortunately, previous studies show strong commonalities in real-world table structures. 
First, tables are arranged in matrices with vertical, horizontal, or both orientations~\cite{nishida2017understanding}. Second, tables usually contain headers on the top or left side to describe other cells~\cite{zanibbi2004survey,wang2016tabular}. Third, headers are often organized with the hierarchical tree structure~\cite{chen2014integrating,chen2013automatic,lim1999automated}, especially in finance and government tables\footnote{SAUS (Statistical Abstract of US from the Census Bureau) is a widely-studied public dataset, in which most tables contain hierarchical headers.}. 
Motivated by these, we propose a \textbf{bi-dimensional coordinate tree}, the first structure to systematically define cell location and cell distance in generally structured tables considering both spatial and hierarchical information. 
Upon this, we propose \textbf{TUTA, a structure-aware method for Table Understanding with Tree-based Attention.} TUTA introduces two mechanisms to utilize structural information.
\textbf{(1)} 
To encode hierarchical information in TUTA, we devised tree-based positional encodings. Different from explicit tree-based positional encodings based on uni-tree structure~\cite{shiv2019novel}, TUTA compares both \textbf{explicit and implicit positional encodings} based on bi-tree structure. And to jointly encode hierarchical and spatial information, TUTA combines tree-based coordinates 
with rectangular Cartesian coordinates. 
\textbf{(2)} 
To enable local cells to effectively aggregate their structural neighboring contexts in large bi-dimensional tables, and ignore unrelated or noisy contexts, we devise a \textbf{structure-aware attention mechanism}. Different from the existing practice of bottom-up tree-based attention~\cite{nguyen2020tree} and constituent tree attention~\cite{wang2019tree} in NL domain, TUTA adapts the general idea of graph attention networks~\cite{velivckovic2017graph} to tree structures to enable both top-down, bottom-up, and peer-to-peer data flow in the bi-tree structure.

Key contributions of this paper are summarized as follows:
\begin{itemize}\setlength{\itemsep}{0.pt }
\setlength{\itemsep}{0pt}
\setlength{\parsep}{0pt}
\setlength{\parskip}{1pt}
	\item For generally structured tables, we devise a bi-dimensional tree to define cell coordinates and cell distance in generally structured tables.
	Based on this bi-tree, we propose TUTA, a structure-aware pre-training method with a distance-aware self-attention. To better incorporate the spatial and structural information, we devise two crucial techniques, called tree position embedding and tree-based attention, that prove to be highly effective throughout experiments. 
	\item We employ three novel pre-training tasks in TUTA, including Masked Language Model at the token level (MLM), multi-choice Cloze at the cell level (CLC), and context retrieval at the table level (TCR). During pre-training, TUTA progressively learns token/cell/table representations on a large volume of tables in an unsupervised manner. 
	\item To demonstrate the effectiveness of TUTA, we fine-tune our pre-trained model on two critical tasks in table structure understanding: Cell Type Classification (CTC) and Table Type Classification (TTC). On both tasks, TUTA is the first transformer-based method ever being applied. It achieves state-of-the-art across five widely-studied datasets.
\end{itemize}

\noindent Our code and pre-trained model will be publicly available\footnote{\url{https://github.com/microsoft/TUTA_table_understanding/}} after internal release and compliance review.

\begin{figure*}[t]
\vspace{-0.8cm}
\begin{center}
\includegraphics[width=6.9in]{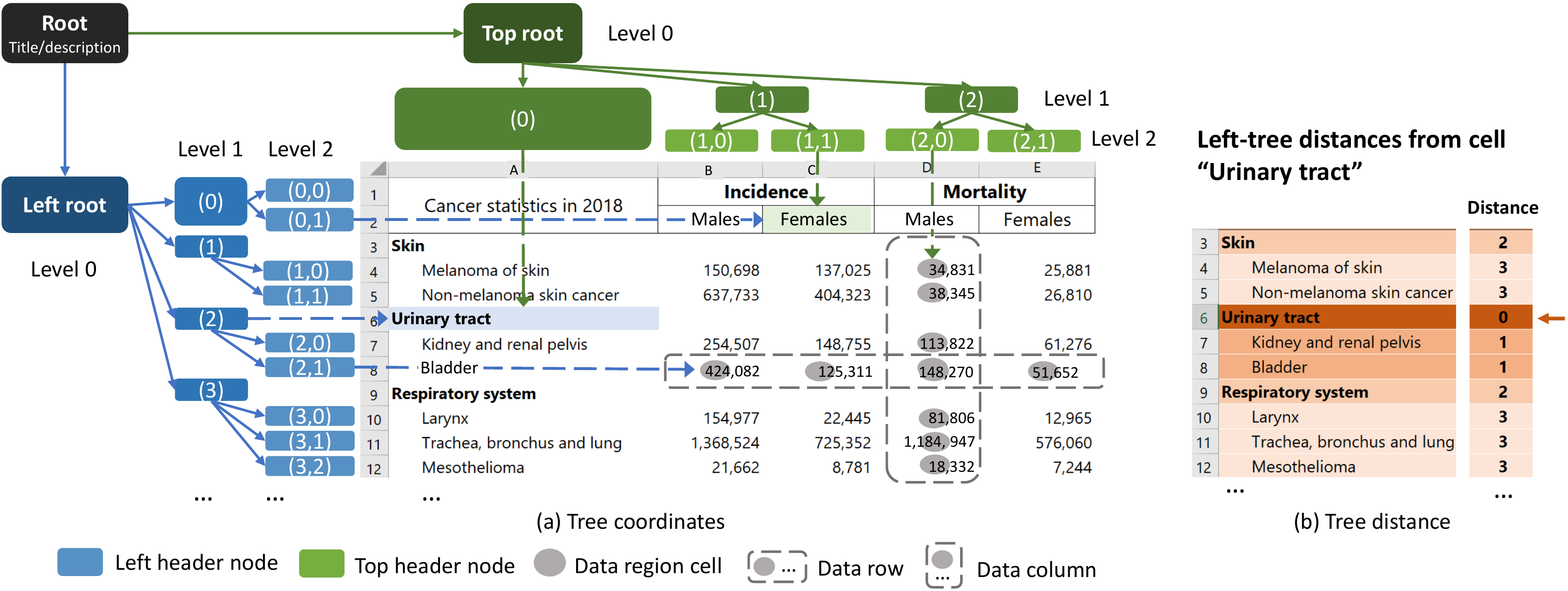}
\end{center}
\vspace{-0.35cm}
\caption{An example to illustrate our proposed tree coordinates and tree distance for generally structured tables. In this example, both the top tree and the left tree contain three levels. Cell ``A6'' (`Urinary tract') is the `parent' node of cells ``A7'' and ``A8'', and has `brothers' including ``A3'' and ``A9''.}
\label{fig:tree}

\end{figure*}

\section{Preliminaries}
\label{sec:pre}

\subsection{Dataset construction}
\label{sec:dataset}

We collect a pre-training corpus of Generally Structured Tables with enlarged volume and diversity in two ways:
(1) \textbf{Structure}: besides relational tables, we also include other structures like entity and matrix tables~\cite{nishida2017understanding}. Extremely small and non-data tables for layout purposes are not used to ensure data quality.
(2) \textbf{Document type}: besides web tables, we also include spreadsheet tables, which are widely used in government, finance, accounting, and medical domains, hence have large quantity and high quality.

We collect \textbf{web tables} from Wikipedia (WikiTable) and the WDC WebTable Corpus~\cite{lehmberg2016large}, and \textbf{spreadsheet tables} from more than $13.5$ million web-crawled spreadsheet files. We end up built a total of \textbf{$57.9$ million} tables, owning much greater volume than datasets used by other table pre-training methods. More details of our data collection, filtering, and pre-processing can be found in Appendix~\ref{app:dataset}.


\subsection{Bi-dimensional coordinate tree}
\label{sec:tree}

As introduced in Section~\ref{sec:intro}, tables usually contain headers on top or left to describe other cells~\cite{zanibbi2004survey,wang2016tabular}. For example, vertical/horizontal tables usually contain top/left headers, and matrix tables usually contain both the top and left headers~\cite{nishida2017understanding}. 
Also, a table is regarded as \textbf{hierarchical} if its header exhibits a multi-layer tree structure~\cite{lim1999automated,chen2014integrating}. 
If not, it reduces to a \textbf{flat} table without any hierarchy. 
To generally model various tables using a unified data structure, we propose a novel bi-dimensional coordinate tree that jointly describes the cell location and header hierarchy.

\noindent \textbf{Tree-based position} \quad  
We define the bi-dimensional coordinate tree to be a directed tree, on which each node has a unique parent and an ordered finite list of children. It has two orthogonal sub-trees: a top tree and a left tree.
For a node in the top/left tree, its position can be got from its path from the top/left tree root.
Each table cell maps to respective nodes on the top and left trees. 
Positions of its top and left nodes combine to a unique \textbf{bi-dimensional tree coordinates}.
As shown in Figure~\ref{fig:tree} (a), the left coordinate of cell ``A6'' is <2> since it's the third child of the left root; successively, ''D8'' is the second child of ``A6'' thus has a left coordinate <2,1>. Using similar methods, ``D8'' has a top coordinate <2,0>.

\noindent \textbf{Tree-based distance} \quad 
In the top/left tree, each pair of nodes connects through a path, or say, a series of steps, with each step either moving up to the parent or going down to a child. 
Motivated by~\cite{shiv2019novel}, we define the \textbf{top/left tree distance} of every node pair to be the step number of the shortest path between them. Then, the \textbf{bi-tree distance} of two cells is the sum of their top and left tree distance.
Figure~\ref{fig:tree} (b) exemplifies distances against cell ``A6'' (`Urinary tract') in the left header. 
\textbf{(1)} Since ``A6'' is the parent node of ``A8'' (`Bladder'), they are only $1$ step apart. 
\textbf{(2)} ``A6'' is the `uncle node' of the cell ``A10'' (`Larynx'), so connecting them goes ``$A6 \rightarrow root \rightarrow A9 \rightarrow A10$'' on the left tree, thus the distance is $3$. 
\textbf{(3)} ``A6'' and ``C2'' (`Females') have an even longer distance of $6$, for it requires moving both on the top ($3$ steps) and left ($3$ steps). 

Note that our bi-dimensional coordinate applies generally on both hierarchical tables and flat tables. \textbf{In flat tables, tree coordinates degenerate to rectangular Cartesian coordinates}, and the distance between two cells in the same row or column is $2$, otherwise is $4$. For texts like titles and surrounding NL descriptions, since they are global information to tables, their distances to cells are set to $0$. This tree-based distance enables effective spatial and hierarchical data-flow via our proposed tree attention in Section~\ref{sec:attention}.

\noindent \textbf{Tree extraction} \quad  
includes two steps: (1) detect header regions in tables and (2) extract hierarchies from headers. 
Methods of header detection have already achieved high accuracy~\cite{fang2012table,dong2019semantic}, we adopt a recent CNN-based method~\cite{dong2019semantic}. 
Next, to extract header hierarchies, we consolidate effective heuristics based on common practices of human formatting, such as the merged cells in the top header and indentation levels in the left header. 
Besides, since formulas containing SUM or AVERAGE are also strong indications of hierarchies, we also incorporate them in this algorithm.
This approach has desirable performance and interpretability when processing large unlabeled corpus. More details can be found in Appendix~\ref{app:tree}.
Since TUTA is a general table pre-training framework based on bi-trees, one can also employ other methods introduced by~\cite{lim1999automated,chen2014integrating,paramonov2020table} to extract headers and hierarchies for TUTA.

\section{TUTA Model}
\label{sec:method}
\begin{figure}[t]
\begin{center}
\includegraphics[width=3.35in]{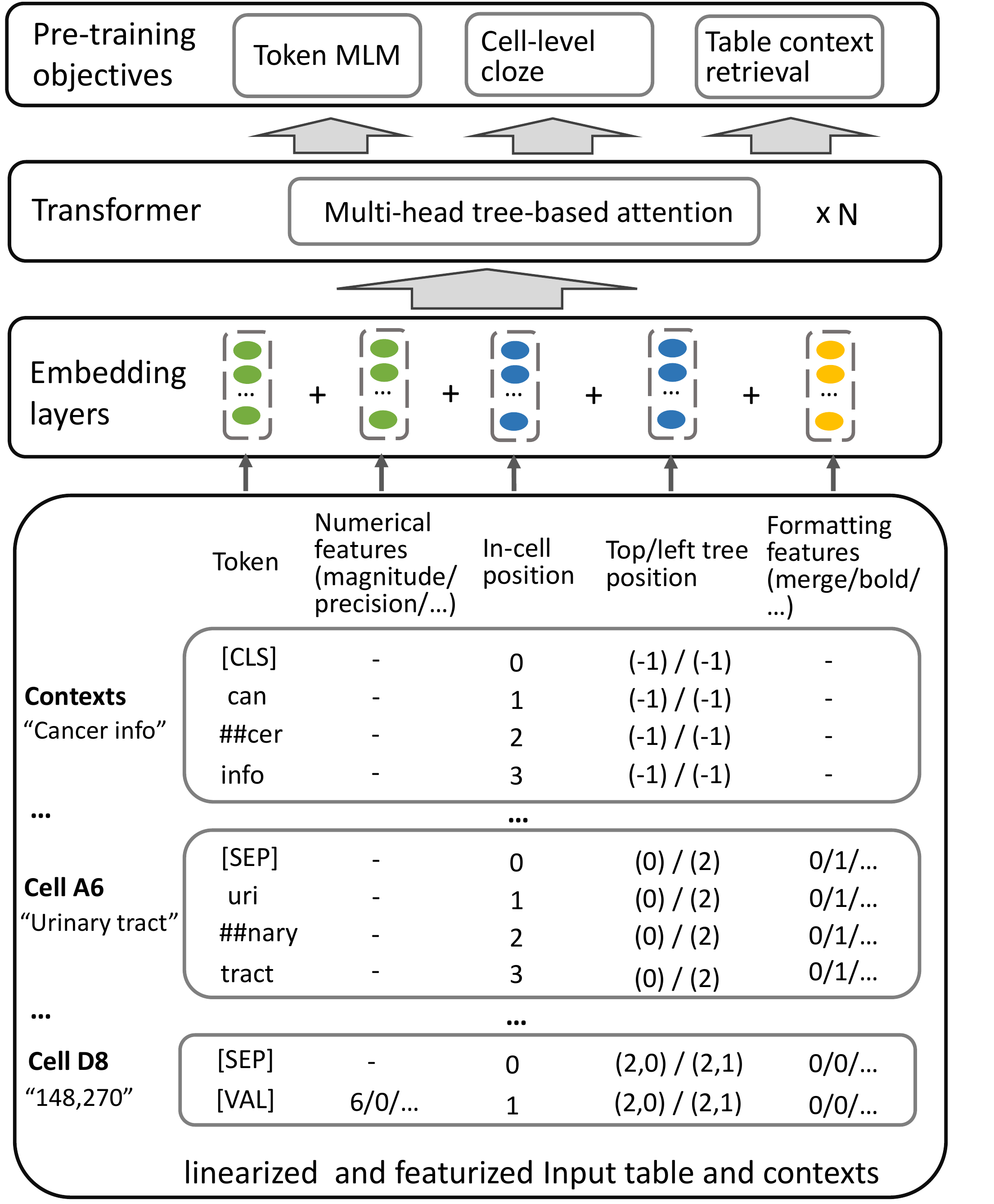}
\end{center}
\caption{ Overview of TUTA architecture. It contains 
(1) embedding layers that convert tables to input embeddings, 
(2) N-layer transformers to capture semantics and structures,
and (3) final projection layers for pre-training objectives. }
\label{fig:architecture}
\end{figure}

Our TUTA model adapts from BERT with four key enhancements: 
(1) build the first dedicated vocabulary on a general source of table corpus to better encode common tokens in real-world tables. 
(2) introduce tree-based position embeddings to incorporate cell location and header hierarchy.
(3) propose a structure-aware attention mechanism to facilitate semantic flows across structurally neighboring contexts.
(4) devise three pre-training objectives to learn progressively the representations at token, cell, and table levels.
The architecture overview is shown in Figure~\ref{fig:architecture}.

\subsection{Vocabulary construction}

Different from long texts in NL documents, cell strings in tables often have short lengths and concise meanings, constituting a word distribution very different from that in NL. For example, to be more compact, tables often record measures `quantity' and  `yards' using abbreviations `qty' and  `yds', or statistical terms `average' and  `difference' as `avg' and `diff'. Hence, directly using NL vocabularies to parse cell strings is improper.

Based on the table corpus introduced in Section~\ref{sec:dataset}, we build the first dedicated vocabulary for tables using the WordPiece model~\cite{devlin2018bert} and get $9,754$ new tokens besides those already in BERT. 
We classify the top-$50$ frequent new tokens in Table~\ref{tab:vocab}. It shows meaningful and intuitive results. Intriguingly, there are indeed plenty of abbreviations, such as `pos', `fg', and `pts' in the sports domain. 
Since NL-form table descriptions like titles also need to be modeled, we merge our table vocabulary with the NL vocabulary~\cite{devlin2018bert} to achieve better generalization, then we perform sub-tokenization for table contexts and cell strings using the Wordpiece tokenizer as in~\cite{devlin2018bert}.

\begin{table}[h]
\begin{center}
    \scalebox{0.88}{
    \begin{tabular}{l  l}
    \whline
    Measures \& units & num (number), qty (quantity), yds, dist, att, eur... \\
    \hline
    Statistical terms & avg (average), pct (percentage), tot, chg, div, rnd... \\
    \hline
    Date \& time & fy (fiscal year), thu (Thursday), the, fri, qtr, yr... \\
    \hline
    Sports & pos (position), fg (field goal), pim, slg, obp... \\
    \hline
     & pld, nd, ast, xxl, px, blk, xxs, ret, lmsc, stl, ef, wkts, \\
    Unsorted & fga, wm, adj, tweet, comp, mdns, ppg, bcs, sog, \\
    & chr, xs, fta, mpg, xxxl, sym, url, msrp, lng \\
    \whline
    \end{tabular}
    }
\caption{ Top-$50$ frequent new tokens in TUTA vocabulary. 
\vspace{-0.3cm}
}\label{tab:vocab}
\end{center}
\end{table}

\subsection{Embedding layer}
Compared with NL text, tables involve much richer information like text, positions, and formats. 
Tokens are jointly embedded by their: in-table position $E_{t\_pos}$, in-cell position $E_{c\_pos}$, token semantics $E_{tok}$, numerical properites $E_{num}$, and formatting features $E_{fmt}$.

\noindent \textbf{In-table position} \quad 
As introduced in Section~\ref{sec:tree}, cells use bi-tree coordinates to store hierarchical information. 
For trees of different depths to get coordinates with unified length, we extend top/left tree coordinates to a predefined max length $L$, as detailed in Appendix~\ref{app:exp-general}. 
Next, we compare two embedding methods for tree positions. 
For one, we adopt the explicit tree embedding in~\cite{shiv2019novel}. It can be explicitly computed and directly extended to our bi-tree structure. For another, we use randomly initialized weights as implicit embeddings and train them jointly with attention layers as~\cite{devlin2018bert}. 

\begin{figure}[t]
    \begin{center}
    \includegraphics[width=3.3in]{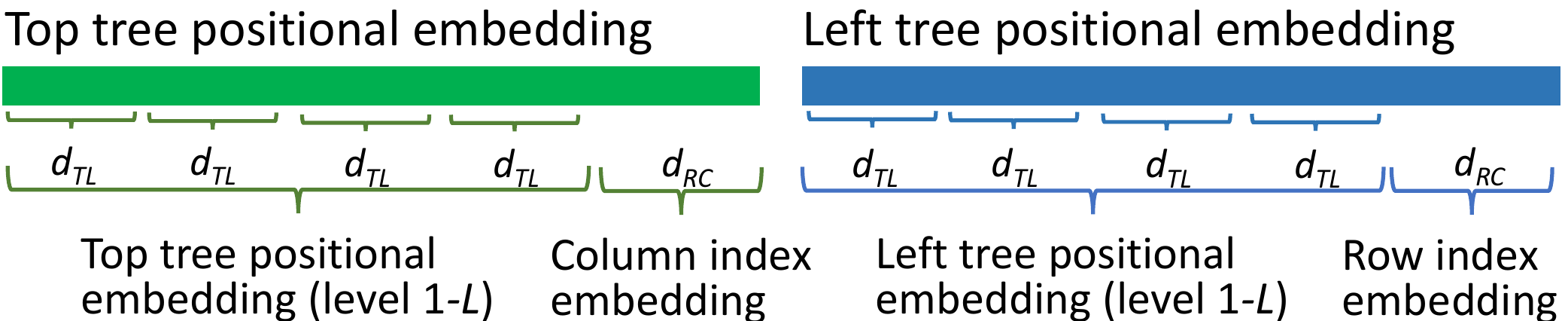}
    \end{center}
    \caption{Size specification of in-table position embeddings. }
    \label{fig:pos_enc}
\end{figure}

To further incorporate spatial information, column and row indexes are combined with the top and left coordinates. 
As shown in Figure~\ref{fig:pos_enc}, we assign each level of top and left coordinates with a sub-embedding (size $d_{TL}$), then concatenate them with row and column sub-embeddings (size $d_{RC}$).
The joint embeddings are then formulated as $E_{t\_pos} = ({W_{t} \cdot x_{t}}) \oplus ({W_{l} \cdot x_{l}}) \oplus ({W_{c} \cdot x_{c}}) \oplus ({W_{r} \cdot x_{r}})$, where $W_{t}, W_{l} \in \mathbb{R}^{d_{TL} \times (\sum_{i=0}^{L-1} G_i)}$ are implicit tree embedding weights, $x_{t}, x_{l} \in \mathbb{R}^{\sum_{i=0}^{L-1} G_i}$ are concatenated $L$-hot vectors for top and left tree coordinates, $W_{r}$, $W_{c}$ $\in  \mathbb{R}^{d_{RC} \times G_{L-1}}$ are row and column embedding weights, $x_{r}, x_{c} \in \mathbb{R}^{G_{L-1}}$ are one-hot vectors for row and column indexes, $G \in \mathbb{N}^{L}$ are maximum node degrees for $L$ tree-layers, and $\oplus$ represents vector concatenation at the last dimension.

\noindent \textbf{In-cell position} \quad  
In-cell position is the index of a token inside a cell. Each position can be simply assigned with a trainable embedding: $E_{c\_pos} = W_{c\_pos} \cdot x_{c\_pos}$, where $W_{c\_pos} \in \mathbb{R}^{H \times I}$ is the learnable weight, $x_{c\_pos}$ is the one-hotted position, and $I$ is predefined to be the maximum number of tokens in a cell.

\noindent \textbf{Token \& Number} \quad  
The token vocabulary is a finite set of size $V = 30,522$, so each token can be simply assigned with a trainable embedding in $W_{tok} \in \mathbb{R}^{H \times V}$. 
However, numbers construct an infinite set, so we extract four discrete features and one-hotly encode them: magnitude ($x_{mag}$), precision ($x_{pre}$), first digit ($x_{fst}$), and last digit ($x_{lst}$). They are embeded using weights $W_{mag}$, $W_{pre}$, $W_{fst}$ and $W_{lst}$, and concatenated at the last dimension: 

\noindent $E_{num} = ({W_{mag} \cdot x_{mag}}) \oplus ({W_{pre} \cdot x_{pre}}) \oplus ({W_{fst} \cdot x_{fst}}) \oplus ({W_{lst} \cdot x_{lst}})$.

\noindent \textbf{Format} \quad  
Formats are very helpful for humans to understand tables, 
so we use format features to signify if a cell has merge, border, formula, font bold, non-white background color, and non-black font color. See a detailed feature list in Appendix~\ref{app:dataset}.
We get the joint formatting embedding by linearly transforming these cell-level features: 
$E_{fmt} = W_{fmt} \cdot x_{fmt} + b_{fmt}$, where $W_{fmt} \in \mathbb{R}^{H \times F}$, $b \in \mathbb{R}^{H}$, and $F$ is the number of formatting features.

\noindent Eventually, the embedding of a token is the sum of all components:
\indent \indent \indent \indent \indent $E = E_{tok} + E_{num} + E_{c\_pos} + E_{t\_pos} + E_{fmt}$

\noindent More detailed specifications of embedding layers can be found in Appendix~\ref{app:exp-model}.

\subsection{Tree attention}
\label{sec:attention}
Based on the bi-tree coordinate and distance (Section~\ref{sec:tree}), we now introduce the tree-based attention. Self-attention is considered to be more parallelizable and efficient than structure-specific models like tree-LSTM and RNN~\cite{shiv2019novel,nguyen2020tree}, but is challenged by lots of `distraction' in large bi-dimensional tables. Because in its general formulation, self-attention allows every token to attend to every other token, without any hints on structure (e.g., if two tokens appear in the same cell, appear in the same row/column, or have a hierarchical relationship). Since spatial and hierarchical information is highly important for local cells to aggregate their structurally related contexts and ignore noisy contexts, we aim to devise a structure-aware  attention mechanism.

Motivated by the general idea of graph attention networks~\cite{velivckovic2017graph}, we inject the tree structure into the attention mechanism by performing masked attention --- using a symmetric binary matrix $M$ to indicate visibility between tokens. 
Given the $i^{th}$ token $tok_i$ in table sequence and its structural neighborhood $SN_i$, we set $M_{i,j} = 1$ if $tok_j \in SN_i$, otherwise $M_{i,j} = 0$.
Based on the tree distance in Section~\ref{sec:tree}, neighbor cells are those within a predefined distance threshold $D$, and neighbor tokens are those of the same or neighboring cells.
Smaller $D$ can enforce local cells to be more ``focus'', but when $D$ is large enough, tree attention works the same with global attention. Succeeding the general attention form of~\cite{vaswani2017attention} using $Q, K, V \in \mathbb{R}^{H \times H}$, our tree-based mask attention simply reads: 
$BiTreeAttenion(Q, K, V) = Attention(Q, K, V) \cdot M$.
Different from existing methods such as bottom-up tree-based attention~\cite{nguyen2020tree} and constituent tree attention~\cite{wang2019tree} in NL domain, our encoder with tree attention enables both top-down, bottom-up and peer-to-peer data flow in the bi-tree structure.
Since TUTA's encoder has $N$ stacked layers with attention depth $D$, each cell can represent information from its neighborhood with nearly arbitrary depth ($N \times D$). In Section~\ref{sec:experiments}, we compare different $D$ values for ablation studies.

\subsection{Pre-training objectives}

\begin{figure}[t]
\begin{center}
\includegraphics[width=3.4in]{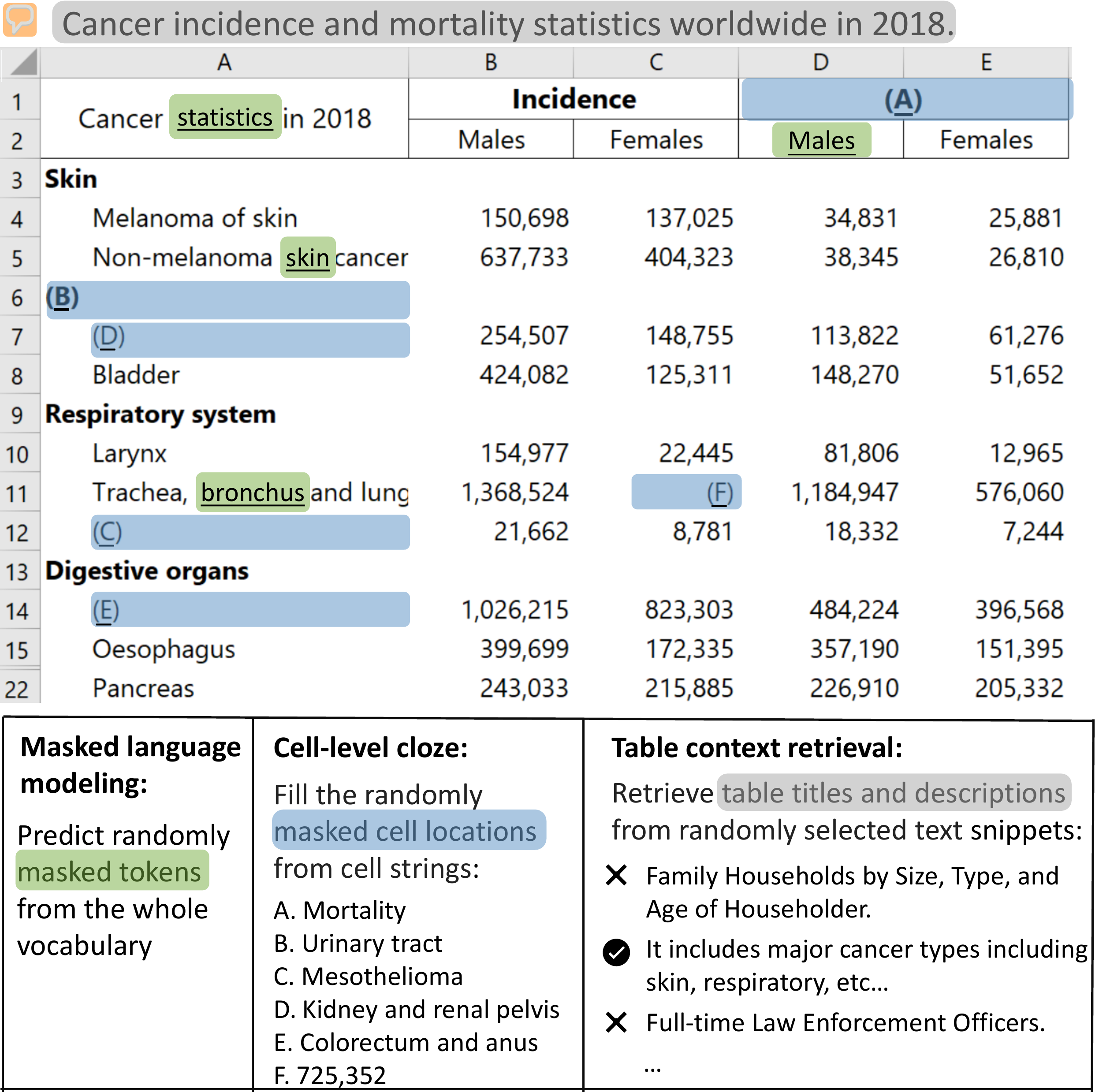}
\end{center}
\caption{ An intuitive example of pre-training objectives. }
\label{fig:task}
\end{figure}

Tables naturally have progressive levels --- token, cell, and table levels. 
To capture table information in such a progressive routine, we devise novel cell- and table-level objectives in addition to the common objective at the token level, as shown in Figure~\ref{fig:task}. 

\noindent \textbf{Masked language modeling (MLM)} \quad 
MLM~\cite{devlin2018bert,lample2019cross} is widely used in NL pre-training. 
Besides context learning within each cell by randomly masking individual tokens, we also use a whole-cell masking strategy to capture relationships of neighboring cells. 
Motivated by~\cite{herzig2020tapas}, we train token representations by predictions on both table cells and text segments. 
MLM is modeled as a multi-classification problem for each masked token with the cross-entropy loss $\mathcal{L}_\text{mlm}$.

\noindent \textbf{Cell-level Cloze (CLC)} \quad 
Cells are basic units to record text, position, and format. So, representations at the cell level are crucial for various tasks such as cell type classification, entity linking, and table question answering. Existing methods~\cite{herzig2020tapas,yin2020tabert,deng2020turl} represent a cell by taking its token average, yet neglect the format and token order. Hence, we design a novel task at the cell-level called Cell-Level Cloze. We randomly
select out cell strings from the table as candidate choices, and at each blanked position, encourage the model to retrieve its corresponding string. As shown in Figure~\ref{fig:task}, one can view it as a cell-level one-to-one mapping from blanked positions to candidate strings.
Given the representations of blanked positions and candidate strings, we compute pair-wise mapping probabilities using a dot-product attention module~\cite{vaswani2017attention}. Next, we perform at each blanked position a multi-classification over candidate strings and compute a cross-entropy loss $\mathcal{L}_\text{clc}$.
Note that, in our attention mechanism, blanked locations can `see' their structural contexts, while candidate strings can only `see' its internal tokens. 

\noindent \textbf{Table context retrieval (TCR)} \quad 
Table-level representation is important in two aspects. On the one hand, all local cells in a table constitute an overall meaning; On the other hand, titles and text descriptions are informative global information for cell understanding. 
So we propose a novel objective to learn table representations. 

Each table is provided with text segments (either positive ones from its own context, or negative ones from irrelevant contexts), from which we use the $[CLS]$ to retrieve the correct ones. 
At the implementation level, given the similarity logit of each text segment against the table representation, TCR is a binary classification problem with cross-entropy loss $\mathcal{L}_\text{tcr}$.

The final objective $\mathcal{L} = \mathcal{L}_\text{mlm} + \mathcal{L}_\text{clc} + \mathcal{L}_\text{tcr}$.

\subsection{Pre-training details} 
\textbf{Data processing} \quad
To get sequential inputs for transformer-based models, we iterate table cells by row and tokenize their strings using the WordPiece Tokenizer. 
Each cell has a special leading token $[SEP]$, while the textual description uses $[CLS]$.
During pre-training, Wiki, WDC, and spreadsheet tables are iterated in parallel to provide diverse table structures and data characteristics.

\noindent \textbf{Model configuration} \quad
TUTA is a $N$-layer Transformer encoder with a configuration aligned with BERT$_{BASE}$. 
To retain BERT's general-semantic representations, TUTA initializes with its token embedding and encoder weights.
TUTA pre-trains on Tesla V100 GPU and consumes about $16M$ tables in $2M$ steps.

Refer more details about data and model to Appendix~\ref{app:exp-model}.

\section{Experiments}
\label{sec:experiments}

Understanding the semantic structures of tables is the initial and critical step for plenty of tasks on tables.
Mainly two tasks receive broad attention: (1) identify the structural type of table cells in the Cell Type Classification (CTC) task; (2) categorize the integrated table structure in Table Type Classification (TTC).

In this section, we first evaluate TUTA on CTC and TTC against competitive baselines. 
Then, we perform ablation studies to show the effectiveness of the critical modules of TUTA.
Last, we analyze typical cases to present more insights on the function of TUTA.

\subsection{Cell Type Classification (CTC)}

To understand table structures, a key step is to identify fine-grained cell types. CTC has been widely studied~\cite{dong2019semantic,gol2019tabular,koci2019deco,gonsior2020active,pujara2021hybrid} with several well-annotated datasets. It requires models to capture both semantic and structure information. Therefore, we use CTC to validate the effectiveness of TUTA.

\begin{figure}[t!]
\begin{center}
\includegraphics[width=3.3in]{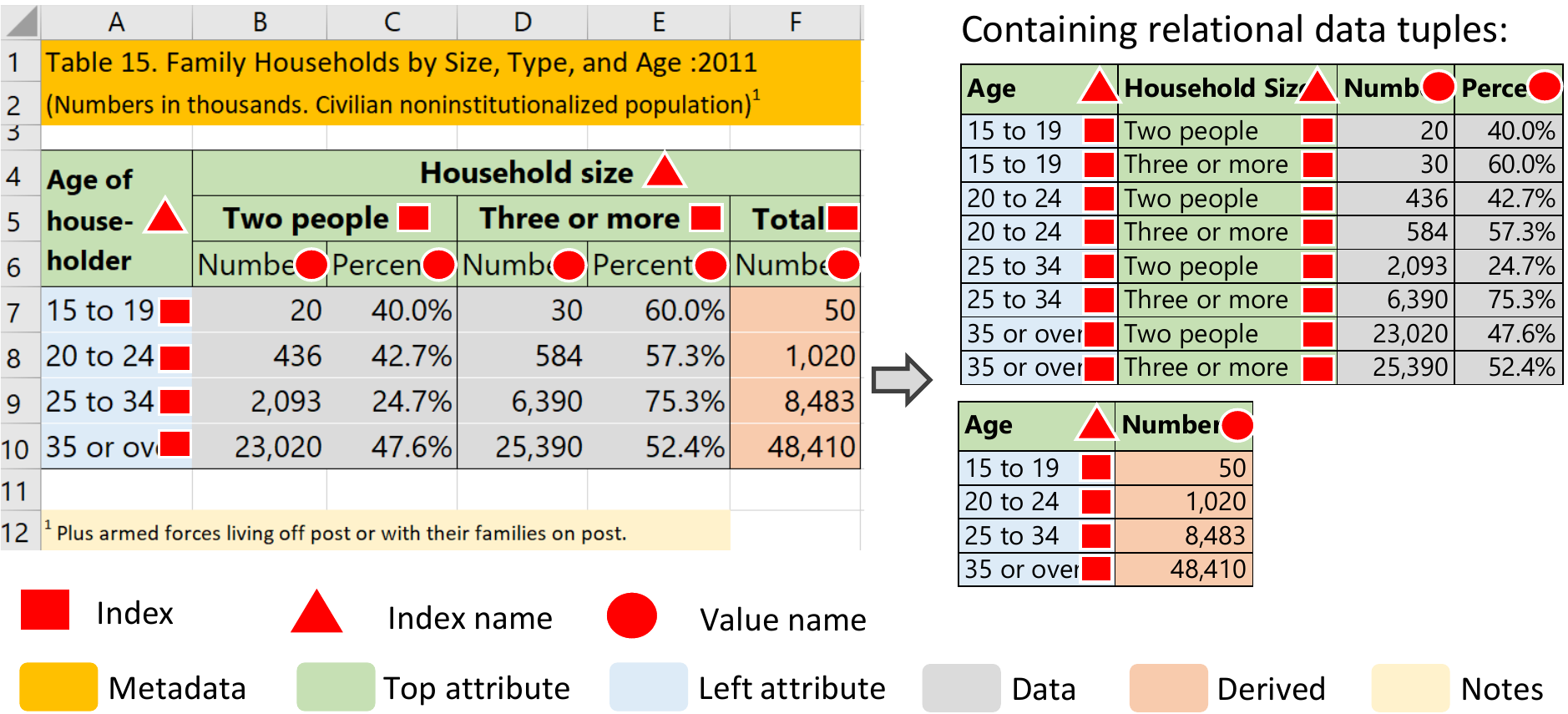}
\end{center}
\caption{ An example to illustrate cell types through relational data extraction. Table on the left is a real case in SAUS (simplified due to space). On the right shows corresponding relational tuples. 
Different cell types (highlighted by different colors and shapes) play different roles in the process of relational data extraction.}
\label{fig:ctc}
\end{figure}

\noindent \textbf{Datasets} \quad 
Existing CTC datasets include WebSheet~\cite{dong2019semantic}, deexcelerator (DeEx)~\cite{koci2019deco}, SAUS~\cite{gol2019tabular}, and CIUS~\cite{gol2019tabular}.
They are collected from different domains (financial, business, agricultural, healthcare, etc.), thus contain tables with various structures and semantics. Table~\ref{tab:ctcdata} shows statistics of table size and structure in each annotated dataset.
Note that they adopt two different taxonomies of cell types. 
DeEx, SAUS, and CIUS categorize cells into general types: metadata (MD), notes (N), data (D), top attribute (TA), left attribute (LA), and derived (B). 
While to enable automatic relational data extraction, WebSheet further defines three fine-grained semantic cell types inside table headers, namely index, index name, and value name~\cite{dong2019semantic}. As Figure~\ref{fig:ctc} shows, cells of different types play different roles in the process of relational data extraction. 

\noindent \textbf{Baselines} \quad 
We compare TUTA with five strong baselines to verify its effectiveness.
CNN$^{BERT}$ ~\cite{dong2019semantic}, Bi-LSTM~\cite{gol2019tabular}, and PSL$^{RF}$~\cite{pujara2021hybrid} are three top existing methods for CTC. 
CNN$^{BERT}$ is a CNN-based method for both cell classification and table range detection using pre-trained BERT. RNN$^{C+S}$ is a bidirectional LSTM-based method for cell classification using pre-trained cell and format embeddings. 
PSL$^{RF}$ is a hybrid neuro-symbolic approach that combines embedded representations with probabilistic constraints.
We also include two representative table pre-training methods, TAPAS~\cite{herzig2020tapas} and TaBERT~\cite{yin2020tabert}.
To ensure an unbiased comparison, we download the pre-trained TAPAS and TaBERT models, and fine-tune them using the same CTC head and loss function with TUTA.
Note that, PSL$^{RF}$ is a relevant yet very recent work evaluated on DeEx, SAUS, and CIUS, so we include it to be a strong baseline. 
However, since the code has not been publicly available, we have not evaluated it on WebSheet.

\begin{table}
\begin{center}
\caption{  Statistics of CTC datasets. }\label{tab:ctcdata}
\scalebox{0.95}{
	\begin{tabular}{l  r  r  r r}
	\whline
	& WebSheet & SAUS & DeEX & CIUS  \\
	\hline
	Number of labeled tables  & 3,503 & 221 &284 & 248 \\
	\hline
	Number of labeled cells  & 1,075k & 192k & 711k & 216k \\
	\hline
	Avg. number of rows & 33.2 & 52.5 & 220.2 & 68.4 \\
	\hline
	Avg. number of columns & 9.9 & 17.7 & 12.7 & 12.7 \\
	\hline
	Prop. of hierarchical tables & 53.7\% & 93.7\% & 43.7\% & 72.1\%  \\
	\hline
	Prop. of hierarchical top & 35.8\% & 68.8\% & 28.9\% & 46.8\% \\
	\hline
	Prop. of hierarchical left & 29.3\%& 76.0\% & 29.2\% & 30.2\% \\
	\whline
    \end{tabular}
}
\end{center}
\end{table}

\noindent \textbf{Fine-tune} \quad 
Tables in CTC datasets are tokenized, embedded, and encoded in the same way as Section~\ref{sec:method}. 
We use the leading $[SEP]$ of cells to classify their types.
Following~\cite{dong2019semantic} on WebSheet and~\cite{gol2019tabular} on DeEx, SAUS, and CIUS, we use the same train/validation/test sets for TUTA and baselines, thus no test tables/cells appear at training. 
Please find more details of data processing and model configuration in Appendix~\ref{app:exp-task}.

\noindent \textbf{Experiment results} \quad 
Results are measured using Macro-F1, for it's a common method to evaluate the overall accuracy over multiple classes. As shown in Table~\ref{tab:ctc}, TUTA achieves an averaged macro-F1 of 88.1\% on four datasets, outperforming all baselines by a large margin (3.8\%+). We observe that RNN$^{C+S}$ also outperforms TaBERT and TAPAS. It is probably because RNN$^{C+S}$ is better at capturing spatial information than TAPAS and TaBERT. TAPAS takes spatial information by encoding only the row and column indexes, which can be insufficient for hierarchical tables with complicated headers. TaBERT, without joint coordinates from bi-dimensions, employs row-wise attention and subsequent column-wise attention. Due to such indirect position encoding, it performs not as well on CTC. 

In Table~\ref{tab:ctc}, we also dig deeper into F1-scores for each cell type in WebSheet, where TUTA consistently achieves the highest score. 
Note that, WebSheet, to perform relational data extraction, further categorizes the header cells into fine-grained types. This is a quite challenging task, for table headers have diverse semantics under complicated hierarchies. Hence, it significantly demonstrates the superiority of TUTA in discriminating fine-grained cell types.

\begin{table}
\begin{center}
\caption{ Results of F1-scores on CTC datasets. IN, I, and VN stand for the IndexName, Index, and ValueName types. }\label{tab:ctc}
\scalebox{0.85}{
    \begin{tabular}{l | c c c c | c c c |c}
    \whline
    \multirow{2}{*}{(\%)} &\multicolumn{4}{c|}{WebSheet}	&\multirow{1}{*}{DeEx}
    &\multirow{1}{*}{SAUS}
    &\multirow{1}{*}{CIUS} &\multirow{1}{*}{Total}\\
    {} & {IN} & {I} & {VN} & \multicolumn{1}{c|}{Macro-} & \multicolumn{3}{c|}{Macro-} & \multicolumn{1}{c}{Avg.}\\
    \whline
    CNN$^{_{^{BERT}}}$ & 69.9 & 86.9 & 78.4 & 78.4& 60.8 & 89.1 & 95.1 & 80.9 \\
    \hline 
    RNN$^{_{^{C+S}}}$ & 75.0 & 86.6 & 77.1 & 79.6 & 70.5 & 89.8 & 97.2 & 84.3 \\
    \hline
    PSL$^{_{^{RF}}}$ & - & - & - & - & 62.7 & 89.4 &  96.7 & - \\
    \hline
    TaBERT$^{_{^{L}}}$ & 76.8 & 85.2 & 75.8 & 79.3 & 50.0 & 78.9 & 92.9 & 75.3 \\
    \hline
    TAPAS$^{_{^{L}}}$ & 74.3 & 88.1 & 84.6 & 82.3 & 68.6 & 83.9 & 94.1 & 82.2 \\
    \hline
    TUTA & \textbf{83.4} & \textbf{91.6} & \textbf{84.8} & \textbf{86.6} & \textbf{76.6} & \textbf{90.2} & \textbf{99.0} & \textbf{88.1} \\
    \whline
    \end{tabular}
}
\end{center}
\end{table}


\subsection{Table Type Classification (TTC)}

In this section, we use a table-level task, TTC, to further validate the effectiveness of TUTA on grasping table-level information. It is a critical task on table structural type classification, and also receives lots of attention~\cite{ghasemi2018tabvec, nishida2017understanding,eberius2015building, lautert2013web, crestan2011web}.

\noindent \textbf{Dataset} \quad
There are various taxonomies for categorizing table types ~\cite{eberius2015building, lautert2013web, crestan2011web}, however, most datasets are not publicly available, except \textsl{the July 2015 Common Crawl} (WCC) in \cite{ghasemi2018tabvec}. Fortunately, this dataset annotates with one of the most general web table taxonomy introduced by \cite{crestan2011web}. Tables are categorized into five types: relational (R), entity (E), matrix (M), list (L), and non-data (ND).

\noindent \textbf{Baselines} \quad
We compare TUTA with five strong baselines on TTC: 
(1) DWTC \cite{eberius2015building}: a Random Forest method based on expert-engineered global and local features,
(2) TabNet \cite{nishida2017understanding}: a hybrid method combining LSTM and CNN, 
(3) TabVec \cite{ghasemi2018tabvec}: an unsupervised clustering method based on expert-engineered table-level encodings. 
Due to the absence of public code and data for DWTC and TabNet, we borrow their evaluation results on WCC from \cite{ghasemi2018tabvec}. 
(4)(5) Similarly to CTC, we also include TAPAS-large and TaBERT-large as two strong baselines.

\noindent \textbf{Fine-tune} \quad
We tokenize, embed, and encode WCC tables as in Section~\ref{sec:method}.
We use $[CLS]$ to do multi-type classification.
Following the training and testing pipeline in \cite{ghasemi2018tabvec}, we use 10-fold cross-validation. 
Read more details about data and parameters in Appendix~\ref{app:exp-task}.

\noindent \textbf{Experiment results} \quad
Table \ref{tab:ttc-result} lists F1-scores on five table types and overall macro-F1 scores.
TUTA outperforms all baselines by a large margin (4.0\%+) on macro-F1.
Comparison results also show some interesting insights:
(1) Although in the CTC task, TaBERT performs worse than Tapas due to neglecting bi-dimensional cell coordinates, in TTC, TaBERT outperforms TAPAS by a large margin. We think the underlying reason is that the sequent row-wise and column-wise attention helps TaBERT better capture global information than TAPAS.
(2) DWTC achieves a high macro-F1 score, but fails to give correct predictions for the List type. After our study, we find that DWTC uses different table type definitions from WCC, especially for the List type, thus the expert-engineered features are too dedicated to achieving desirable results on WCC. In contrast, representation learning methods are more transferable from one taxonomy to another. 
(3) TabNet fails to get reasonable results on most table types~\cite{ghasemi2018tabvec}. We believe the reason is that the model size of TabNet (a combination of CNN and RNN) is too large to perform well on tiny datasets (WCC has $386$ annotated tables in total). In contrast, TUTA, TAPAS, and TaBERT can generalize well to downstream tasks via self-supervised pre-training.

\begin{table}
	\begin{center}
	\caption{ Evaluation results on WCC. R, E, M, L, ND stand for relational, entity, matrix, list, and non-data table types. }\label{tab:ttc-result}
	\begin{tabular}{l  c c c c c | c}
	\whline
	(\%) & R & E & M & L & ND & Macro-F1 \\
	\whline
	
	DWTC  & 83.0 & \textbf{91.0} & 73.0 & 0.0 & 79.0 & 79.0 \\
	\hline
	TabVec  & 70.0 & 53.0 & 47.0 & 55.0 & 71.0 & 63.0 \\
	\hline
	TabNet  & 9.0 & 41.0 & 9.0 & 11.0 & 25.0 & 27.0 \\
	\hline 
	TAPAS$^{_{^{L}}}$ & 73.3 & 77.4 & \textbf{90.5} & 80.5 & 78.5 & 80.0 \\
	\hline
	TaBERT$^{_{^{L}}}$ & 87.3 & 87.5 & 70.9 & 86.0 & 86.2 & 83.6 \\
	\hline
	TUTA & \textbf{88.7} & 77.9 & 81.7 & \textbf{100.0} & \textbf{89.9} & \textbf{87.6}\\
	\whline
	\end{tabular}
	\end{center}
\end{table}


\subsection{Ablation studies}
\label{sec:ablation}
To validate the effectiveness of Tree-based Attention (TA), Position Embeddings (PE), and three pre-training objectives, we evaluate eight variants of TUTA. 

We start with \textit{TUTA-base} before using position embeddings to prove the function of using and varying TA distances. One variant without TA and three of visible distances $8, 4, 2$ are tested.

\begin{itemize}\setlength{\itemsep}{0.pt }
    \setlength{\itemsep}{0pt}
    \setlength{\parsep}{0pt}
    \setlength{\parskip}{1pt}
    \item \textit{TUTA-base, w/o TA}: cells are globally visible.
    \item \textit{TUTA-base, TA-8/4/2}: cells are visible in a distance of 8/4/2.
\end{itemize}

Upon this, we keep the distance threshold $D = 2$ and augment TUTA-base with explicit or implicit positional embeddings.

\begin{itemize}\setlength{\itemsep}{0.pt }
    \setlength{\itemsep}{0pt}
    \setlength{\parsep}{0pt}
    \setlength{\parskip}{1pt}
    \item \textit{TUTA-implicit}: embed positions using trainable weights.
    \item \textit{TUTA-explicit}: compute positional inputs explicitly as in~\cite{shiv2019novel}.
\end{itemize}

Further, to measure the contribution of each objective, we respectively remove each from the \textit{TUTA-implicit} (the best so far).

\begin{itemize}\setlength{\itemsep}{0.pt }
    \setlength{\itemsep}{0pt}
    \setlength{\parsep}{0pt}
    \setlength{\parskip}{1pt}
    \item \textit{TUTA, w/o MLM}; \textit{TUTA, w/o CLC}; \textit{TUTA, w/o TCR}
\end{itemize}

\begin{table}
\caption{  Ablation results on CTC and TTC datasets.}\label{tab:abl-both}
\scalebox{0.89}{
    \begin{tabular}{l | c c c c c | c}
    \whline
    \multirow{2}{*}{Macro-F1 scores (\%)} &
    \multicolumn{5}{c|}{CTC} & \multicolumn{1}{c}{TTC}\\
     &\multicolumn{1}{c}{WebSheet}	&\multirow{1}{*}{DeEx}
    &\multirow{1}{*}{SAUS}
    &\multirow{1}{*}{CIUS} &\multirow{1}{*}{Avg.} 
    &\multirow{1}{*}{WCC}\\
    
    \hline
    TUTA-base, w/o TA &  76.3 & 70.5  &  80.0 & 92.8 & 79.9 & 77.4 \\
    TUTA-base, TA-8 & 80.0 & 71.1 & 80.5 & 93.2 & 81.2 & 78.0  \\
    TUTA-base, TA-4 & 81.5 & 73.8 & 80.9 & 95.6 & 83.0 & 81.4  \\
    TUTA-base, TA-2 & 84.5 & 75.5 & 84.6 & 96.7 & 85.3  & 82.4  \\
    \hline
    TUTA-explicit  & 86.5 &  76.0 &  89.7 &  98.8 & 87.8 & 85.0 \\
    TUTA-implicit  & \textbf{86.6} & \textbf{76.6} & \textbf{90.2} & \textbf{99.0}  & \textbf{88.1} & \textbf{87.6} \\
    \hline
    TUTA, w/o MLM	& 85.4 & \textbf{76.6} & 89.2 & \textbf{99.0} & 87.6 & 84.0 \\
    TUTA, w/o CLC	& 83.0 & 76.4 & 88.7 & 98.9 & 86.8 & 82.0 \\
    TUTA, w/o TCR	& 85.8 & \textbf{76.6} & 88.2 & \textbf{99.0}  & 87.4 & 81.8 \\
    \whline
    \end{tabular}
}
\end{table}

\noindent \textbf{Experiment results}  \quad 
Table~\ref{tab:abl-both} shows the ablation results on CTC and TTC.
It is clear that smaller attention distances help \textit{TUTA-base} to achieve better results. In cell type classification, \textit{TUTA-base} can only get 79.9\% of averaged macro-F1, lower than 82.2\% achieved by TAPAS.
But as the distance threshold decreases to 2, \textit{TUTA-base,TA-2} improves a lot (5.4\%), then outperforms TAPAS by 3.1\%.

Further, tree position embeddings improve accuracy as well.
In cell type classification, \textit{TUTA-implicit} gets an $88.1\%$ macro-F1 on average, which is $2.7\%$ higher than its base variant \textit{TA-2}, and $0.3\%$ higher than \textit{TUTA-explicit}. 
In table type classification, \textit{TUTA-implicit} also outperforms its base variant and \textit{TUTA-explicit}.
We infer that, encoding spatial and hierarchical information can significantly improve the accuracy of TUTA.

Moreover, removing any pre-training objectives leads to accuracy drops in both tasks.
In the task of cell type classification, while removing MLM or TCR decreases the average macro-F1 by only around $0.5\%$, removing CLC leads to a greater drop of $1.3\%$, showing the importance of CLC in this cell-level task.  
Note that for SAUS, removing TCR also causes an obvious accuracy drop (2\%).
We find that most tables in SAUS have informative titles and descriptions, which provide strong hints to enhance local cell understanding.
In the task of table type classification, removing MLM causes a $3.6\%$ drop, while removing CLC or TCR drops even more by around $6\%$. 

\subsection{Case studies}
\label{sec:case}
To facilitate an intuitive understanding, we show two typical cases in the CTC test set to help illustrate the experiment results.

\begin{figure}
    \centering
    \caption{A case in WebSheet, TUTA well captures the fine-grained cell types in headers.}
    \includegraphics[width=8.5cm]{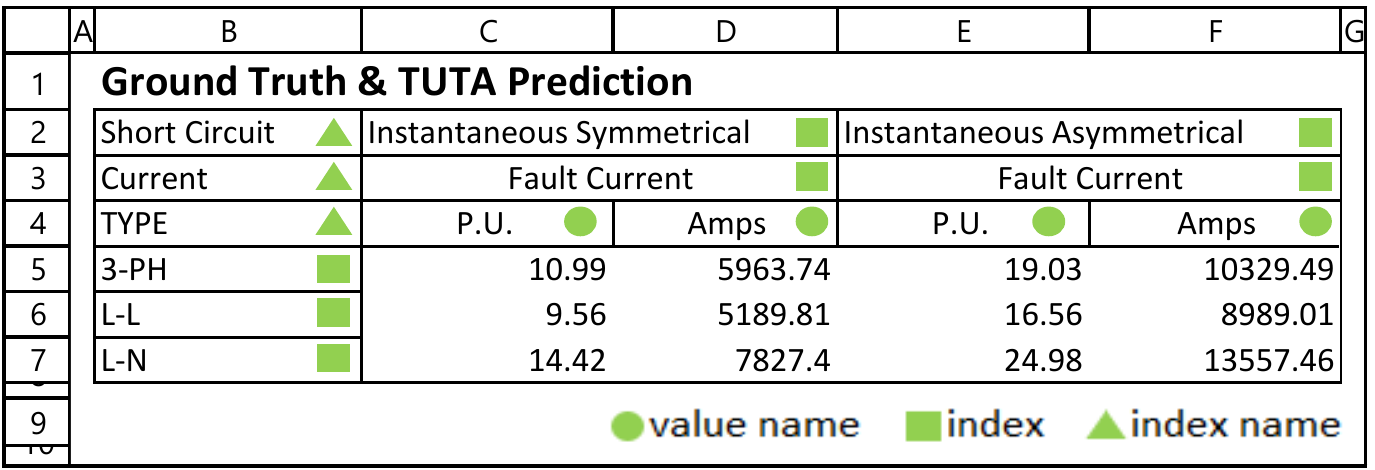}
    \label{fig:case-ctctypes}
\end{figure}

Figure~\ref{fig:case-ctctypes} it TUTA prediction on a WebSheet table, where header cells have complex structural and semantic relationships. Though this is a fairly challenging case, TUTA correctly predict all of them.

\begin{figure}
    \centering
    \caption{In a case of SAUS, TUTA associates the structure and semantics between header and data.}
    \includegraphics[width=8.5cm]{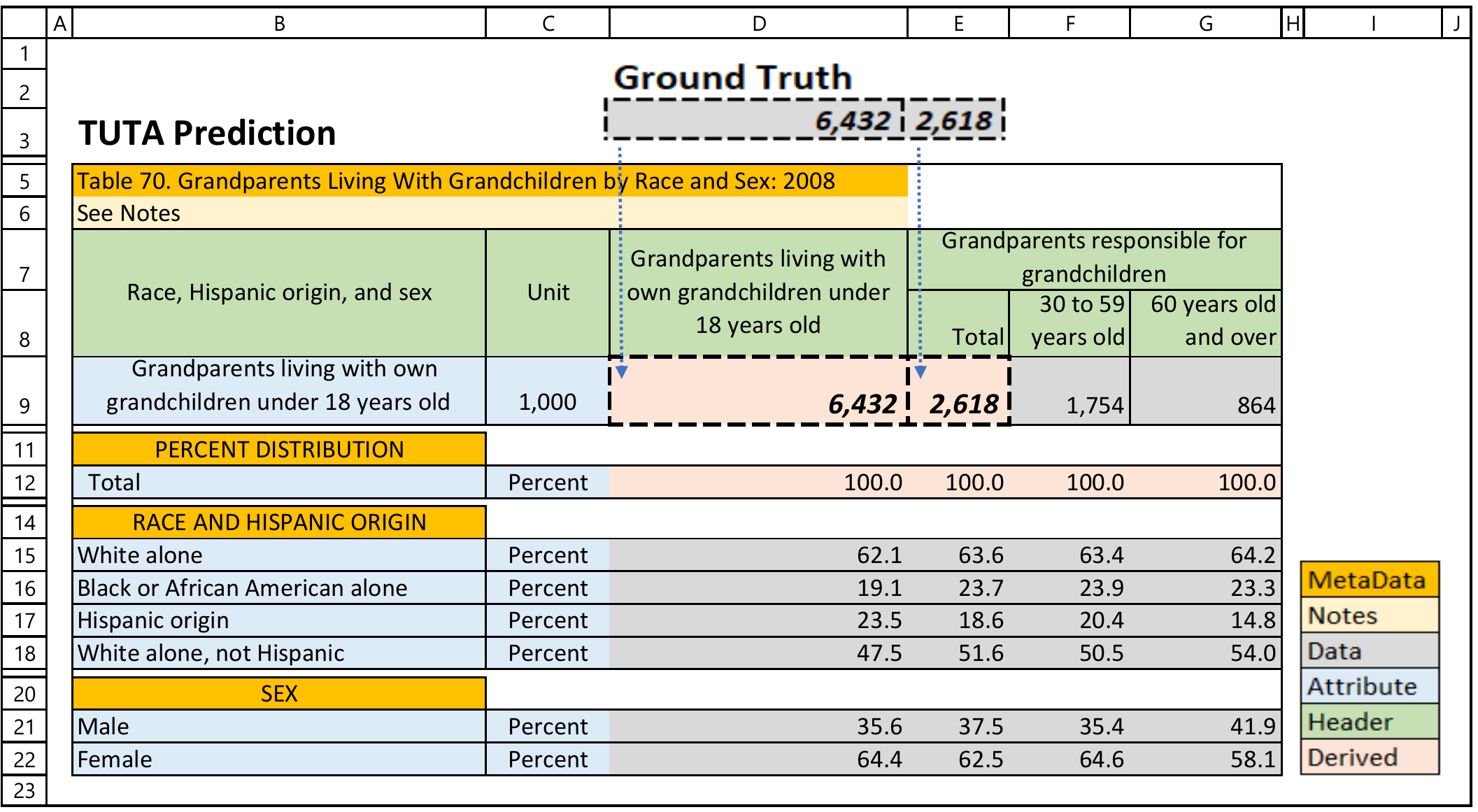}
    \label{fig:case-derived}
\end{figure}

Figure~\ref{fig:case-derived} shows a bad case in SAUS, where ``D9'' and ``E9'' are labeled to the `Data' type, yet TUTA classifies them to `Derived' (a derivation of other `Data' cells). 
After our study, we find that both the TUTA's prediction and the ground truth are wrong --- ``D9'' belongs to the `Data' type, but ``E9'' should be `Derived'.

For one, ``D9'' looks like to be a `Derived' type since its parent, ``D8'', is merged on the top. Also, ``D8'' is a direct child of the top-tree root, thus has a higher tree level than cells like ``F8'' and  ``G8''. 
These formatting and structural information easily leads the model to view ``D9'' as `Derived'. However, only by iteratively performing possible calculations over data can we know that, ``D9'' is not a derivation. Though numerical computations exceed our current scope of table understanding in this paper, we do think it reasonable to augment numerical computations in future works.

For another, ``E9'' is correctly predicted by TUTA. It is a \textbf{ground truth error caused by human labelers}. It should be a `Derived' cell for it's indeed the sum of ``F9:G9'', even though there is no formula to indicate it. 
We think TUTA makes the correct prediction out of two factors: 
(1) semantics of the keyword `Total' in its parent cell ``E8'', 
(2) the large magnitude of number in ``E9''. 
It demonstrates that TUTA has a fair ability to capture both the structural and semantic information when classifying cells into types.

\section{Related Work}
\label{sec:related}

\noindent \textbf{Table representation learning}\quad 
Though TUTA is the first effort to pre-train transformers on variously structured tables, a range of prior work has also pre-trained transformers on relational tables.
Table-BERT linearized tables as sentences so that tables can be directly processed by the pre-trained BERT model~\cite{chen2019tabfact}. 
TAPAS and TaBERT target question answering over relational tables via joint pre-training of tables and their text~\cite{herzig2020tapas,yin2020tabert}. 
TURL learns representations from relational tables to enhance table knowledge matching and table augmentation~\cite{deng2020turl}, but in TURL, each cell only aggregates information from the located row and column. It’s worth mentioning that, in addition to transformers, prior work has also explored other models. For example,~\cite{gol2019tabular} uses continuous bag-of-words and skip-gram to learn cell embeddings; 
Table2Vec~\cite{zhang2019table2vec} adopts skip-gram neural network models to train word embeddings; and TabNet learns token embeddings jointly with the LSTM model~\cite{nishida2017understanding}.

\noindent \textbf{Neural networks for table understanding}\quad
Since both semantic and spatial information is crucial for table understanding, lots of neural architectures have been proposed to capture spatial and semantic information jointly. CNNs~\cite{dong2019semantic,chen2019colnet} are adopted to capture spatial information for web and spreadsheet tables. Bidirectional RNNs and LSTMs are widely adopted in web tables to capture the order of rows and columns~\cite{nishida2017understanding,gol2019tabular,fetahu2019tablenet,kardas2020axcell}. Later work proposed a hybrid neural network by combining bidirectional RNNs and CNNs in the task of column type prediction~\cite{chen2019learning}. 
Recently, graph neural networks have been explored for table understanding and question answering~\cite{zayats2021representations,zhang2020graph,koci2018table}.
\cite{dong2020neural} use conditional GANs to recommend good-looking formatting for tables.
TUTA is the first transformer-based method for general table structure understanding and achieves state-of-the-arts on two representative tasks.

\section{Conclusion and Discussion}
\label{sec:conclusion}

In this paper, we propose TUTA, a novel structure-aware pre-training model to understand generally structured tables.
TUTA is the first transformer-based method for table structure understanding. It employs two core techniques to capture spatial and hierarchical information in tables: tree attention and tree position embeddings.
Moreover, we devise three pre-training objectives to enable representation learning at token, cell and table levels.
TUTA shows a large margin of improvements over all baselines on five widely-studied datasets in downstream tasks.

TUTA is believed to be effective in other tasks like table Question Answering (QA) and table entity linking as well. 
Since current datasets of these tasks only consider relational tables, we aim to extend the scope of them to variously structured tables later on. 
Gladly, TUTA's superior performance on CTC and TTC soundly proves the effectiveness of structure-aware pre-training.
\bibliographystyle{named}
\bibliography{main}

\appendix
\clearpage

\section{Dataset construction}\label{app:dataset}

\noindent \textbf{Data collection} \quad
Two corpus of tables are used at pre-training.

\textbf{Web tables} \quad 
We collect $2.62$ million web tables from WikiTable~\footnote{https://github.com/bfetahu/wiki\_tables} and 50.82 million web tables from WDC WebTable Corpus~\cite{lehmberg2016large}. We also keep their titles, captions, and NL contexts.

\textbf{Spreadsheet tables} \quad 
Spreadsheets are widely used to organize tables\footnote{By estimation, there are around $800$ million spreadsheet users. https://medium.com/grid-spreadsheets-run-the-world/excel-vs-google-sheets-usage-nature-and-numbers-9dfa5d1cadbd}.
We crawl about $13.5$ million public spreadsheet files (.xls and .xlsx) from more than $1.75$ million web sites. Then we utilize the techniques of TableSense~\cite{dong2019tablesense} to detect tables from sheets of each file. In total, we get about $115$ million tables.


\noindent \textbf{Pre-processing}\quad 
Since spreadsheets are crawled from various web sites, the tables detected from them are very noisy. We hence perform data cleansing and language filter to build a clean table corpus for pre-training.
First, we filter out tables having extreme table size (number of rows/columns $< 4$, number of rows $> 512$, or number of columns $> 128$), tables with over-deep hierarchical headers (number of top/left header rows/columns $> 5$) and tables without any headers. With these rules, we filter $52.34\%$ tables from the original dataset. 
Second, we de-duplicate the filtered data based on table content. After removing the duplicated tables, $23.49\%$ tables are left. 
Third, we use a text analytics tool\footnote{https://azure.microsoft.com/en-us/services/cognitive-services/text-analytics/} for language detection. Among all filtered and de-duplicated tables, about $31.76\%$ tables are English and used for pre-training. Finally, we get $4.49$ million spreadsheet tables for TUTA.

\begin{table}[h]
\begin{center}
\caption{ Comparison of datasets for table pre-training.}\label{tab:data1}
\scalebox{1}{
\begin{tabular}{l  l  l  r}
\whline
&  Data sources & Table types & Total amount \\
\hline
\multirow{3}{*}{TUTA} & WikiTable & \multirow{3}{*}{General}  & 2.6 million  \\
& WDC  &  & 50.8 million \\
& Spreadsheet  &  & 4.5 million\\
\hline
TAPAS  &WikiTable & Relational & 6.2 million \\
\hline
\multirow{2}{*}{TaBERT} & WikiTable & \multirow{2}{*}{Relational}  & 1.3 million  \\
  & WDC & & 25.3 million \\
\hline
TURL & WikiTable & Relational & 0.6 million \\

\whline
\end{tabular}
}
\end{center}
\end{table}


\noindent \textbf{Feature extraction}\quad 
We use ClosedXML\footnote{https://github.com/ClosedXML/ClosedXML} to parse spreadsheet files and extract features. For the two web table corpus, tables are serialized to JSON formats, so we just load and parse the JSON files for feature extraction. We unify the featurization schema for web tables and spreadsheet tables as shown in Table~\ref{tab:feat}.

\begin{table}[h]
\centering
\caption{Feature set of table cells.}\label{tab:feat}
\scalebox{0.9}{
\begin{tabular}{l l c}
\whline

\textbf{Description} & \textbf{Feature value} & \textbf{Default } \\
\hline
\multicolumn{1}{l}{ \textbf{Cell text}} & \multicolumn{1}{l}{ } \\
Cell text & token/number list & - \\
\hline
\multicolumn{1}{l}{ \textbf{Cell position}} & \multicolumn{1}{l}{ } \\
Row/column indexes & a pair of integers & - \\
Tree-based coordinates & a pair of int tuples & - \\
\hline
\multicolumn{2}{l}{ \textbf{Merged region}} \\

 The number of merged rows  & positive integer & 1 \\
 The number of merged columns  & positive integer & 1 \\
\hline
\multicolumn{2}{l}{ \textbf{Data type}} \\

 If cell string matches a date template &   0 or 1  & 0 \\
 If formula exists in the cell &   0 or 1  & 0 \\
\hline
\multicolumn{2}{l} {\textbf{Cell format}} \\

 If the bold font is applied  &   0 or 1  & 0\\
 If the background color is white  &   0 or 1  & 1\\
 If the font color is black  &   0 or 1  & 1\\
\hline
\multicolumn{2}{l}{ \textbf{Cell border}} \\

If cell has a top border &   0 or 1 & 0 \\
If cell has a bottom border &   0 or 1 & 0 \\
If cell has a left border &   0 or 1 & 0 \\
If cell has a right border &   0 or 1 & 0 \\
\whline
\end{tabular}
}
\end{table}

\section{Tree extraction}\label{app:tree}
Given a detected (top or left) header region by~\cite{dong2019semantic}, we develop a rule-based strategy to extract tree hierarchies out of merged cells, indentation levels, and formulas in data cells. By applying this method on the detected top and left header for each table, we get respective tree hierarchies, then build upon them the bi-dimensional coordinate tree designed for TUTA.

\noindent \textbf{Merged cells}\quad 
Merged regions provide spatial alignment and grouping information between levels of headers, thus are useful for finding hierarchical relationships. Specifically, cells covered beneath (or on the right side of) a merged region are often children nodes of the merged parent. Through this inter-level association (usually on the top), we can build a prototype of the hierarchical header tree.

\noindent \textbf{Indentation levels}\quad 
Indentation usually appears in left headers to indicate hierarchical levels in both spreadsheet and web tables. 
Generally, indentation refers to the visual indentation effect that includes various operation methods. Users can insert various number of spaces and/or tabs to create an indentation effect. Or, they can write cells at different levels into separate columns. In spreadsheets, Excel provides an operation to set indentation level in the cell format menu. 
We transform these layout operations into effective heuristics, to extract tree hierarchies from left headers. 
By extracting indentation levels based on these three operations, we accordingly assign different hierarchical levels to cells and create the left header tree.

\noindent \textbf{Formulas}\quad 
Formula is an important feature, especially in spreadsheet tables that contain information about the calculation relationship between cells. Some formulas indicate the aggregation relationship between cells, for example, SUM and AVERAGE. What's more, if all cells in a row share the same aggregation formula pattern, the left header node leading this row should be treated as the parent node of other referenced rows. And the columns work similarly. Then, we can get the hierarchical relationship based on formulas. It has a higher priority than indentation levels in our extracting method.

\section{Experiment details}\label{app:exp}

\subsection{Data pre-processing}\label{app:exp-general}
\noindent \textbf{General pre-processing} \quad
In order to control sequences within a reasonable length and prevent unhelpful numerical information, we: 
(1) perform a heuristic sampling over cells in the data region, for they often express similar numerics yet introduce limited semantics. By classifying data cells into text- and value-dominant types, we sample out $50\%$ in the former and $90\%$ in the latter. 
(2) take at most $8$ tokens from a cell and $64$ tokens from a text segment, given empirical studies showing that ~99\% of the cases meet this condition.

\noindent \textbf{Masked Language Model (MLM)} \quad
When selecting tokens to be masked, we employ a hybrid strategy of token-wise and whole-cell masking. To be more specific, we first randomly select 15\% of the table cells, then for each cell: (1) with a $70\%$ probability, mask one random token, or (2) with a $30\%$ chance, mask all of the tokens of that cell.

\noindent \textbf{Cell-Level Cloze (CLC)} \quad
When choosing cells to be dug out, we assign a higher probability for cells in table headers than those in the data region. Meanwhile, for more attentive training on the structural information, we apply two strategies at cell selection. In one way, we randomly choose cells on the same sub-tree based on our bi-tree structure; in another, cells are randomly taken from different top header rows and left header columns. 
In all, around $20\%$ of the cells are selected in each table.

\noindent \textbf{Table Context Retrieval (TCR)} \quad
We consider table titles and surrounding descriptions to be relevant texts of tables. We split these texts into segments, then provide each table with both positive (from its own context) and negative (from other's context) ones. Each table has at most three positive segments and three negative segments. 
We randomly choose one positive segment and put it after the leading token $[CLS]$. Other segments are assigned with leading $[SEP]$ tokens and padded after the sequence of table tokens.
Note that in our tree-based attention, $[CLS]$ learns table-level representations and can `see' all table cells, while the $[SEP]$ of each text segment learns text segment representations and only `see's its internal tokens. 
Please find more details in our codebase about computations of the attention visibility matrix~\footnote{https://github.com/microsoft/TUTA\_table\_understanding/tuta/model/backbones.py}.


\subsection{Model Configuration}\label{app:exp-model}
\noindent \textbf{Embedding details} \quad
Here we specify the embedding modules.

For in-table positions, we empirically set an upper-bound for table size and header shape. 
To control the maximum supported tree size, we define a maximum number of tree layers $L = 4$, and at deeper levels $i \in [0, L]$, an increasing number of degrees from $32$, $32$, $64$, to $256$, thus a maximum number of node degrees $G \in \mathbb{N}^{L}$ for tree layers. 
In another word, a top/left header has at most $L = 4$ rows/columns; while within the $i^{th}$ row/column, there are at most $G_{i}$ cells, where $G = [32, 32, 64,256], i \in [1, 4]$. 
Note that the last supporting degree, $G_{L-1} = 256$, accords with both the maximum number of rows and columns, allowing a compatible conversion from a hierarchical coordinate to a flat setting.
When encountering large tables, we can split them into several smaller ones (share the same top or left header) based on detected table headers.
When transforming in-table position to embeddings, each index of the tree coordinate is mapped to a dimension of size $d_{TL} = 72$, while row and column indices are embedded to size $d_{RC} = 96$.

For internal position embedding, we keep the first $I = 8$ tokens in each cell and $I = 64$ for each text segment, given the empirical studies on our corpus showing that over $99\%$ of cell strings and texts satisfy this condition. Note that though few ($<1\%$), it is necessary to trim those length-exceeding cases, for a string too long often introduces noise and produces computational inefficiency. 

To embed numbers, magnitude $x_{mag} \in [0,M]$, precision$x_{pre} \in [0, P]$, the first digit $x_{fst} \in [0, F]$, and the last digit $x_{lst} \in [0, L]$, where $M, P, F, L = 10$. Correspondingly, their embedding weights are $W_{mag}, W_{pre}, W_{fst}, W_{lst} \in \mathbb{R}^{M/P/F/L \times \frac{H}{4}}$, hidden size $H = 768$.

For format embeddings, the $F = 11$ integer cell features $x \in \mathbb{N}^{F}$ (as listed in Table~\ref{tab:data1}) undergoes a transformation $E_{fmt} = W_{fmt} \cdot x + b$ by weight $W_{fmt} \in \mathbb{R}^{F \times H}$ and bias $b \in \mathbb{R}^{H}$.

\noindent \textbf{Pre-training hyper-parameters} \quad
Pre-training TUTA has two stages. First, we use table sequences with no more than $256$ tokens with a batch size of $12$ for the first 1M steps of training. Then, we extend the supporting sequence length to $512$ and continue the training for another 1M steps using a batch size of $4$.
The above procedures are implemented with PyTorch distributed training. We train TUTA and its variants on 64 Tesla V100 GPUs, and each TUTA variant takes nine days on four Tesla V100 GPUs.


\subsection{Downstream tasks}\label{app:exp-task}
\noindent \textbf{Cell Type Classification (CTC)} \quad

We use a fine-tune head for this multi-classification problem.
Given the encoder's outputs, we use the leading $[SEP]$ in each cell to represent the whole cell, then perform linear transformations to compute logits over cell types: $p_c = W_2 \cdot~gelu(W_1 \cdot r_c + b_1) + b_2$, where $r_c \in \mathbb{R}^{H}$ is the cell representation, $W_1 \in \mathbb{R}^{H \times H}$ and  $W_2 \in \mathbb{R}^{C \times H}$ are weights, $b_1 \in \mathbb{R}^{H}$ and $b_2 \in \mathbb{R}^{C}$ are bias, and $C$ is the number of cell types. We compute the cross-entropy loss against labels.

When splitting datasets into train, validation, and test, we adopt a table-wise, rather than a cell-wise manner. Since cells in a table are always considered together, none of the cells in test tables have been used for training.

At fine-tuning, we follow~\cite{dong2019semantic} and tune our model for $4$ epochs on WebSheet. 
For DeEx, SAUS and CIUS, since the number of tables is not as big, we separately tune TUTA for $100$ epochs on five random-split folds and report their average macro f1. 
All experiments set the batch size to $4$ and the learning rate to $8\mathrm{e}{-6}$.

\noindent \textbf{Table Type Classification (TTC)} \quad
Similar to CTC, we use a dedicated head to classify table sequences into types.
Taking the leading $[CLS]$ of each table to represent the whole table, we compute logics over table types: $p_t = W_2 \cdot~gelu(W_1 \cdot r_t + b_1) + b_2$, where the table representation $r_t \in \mathbb{R}^{H}$, weights $W_1 \in \mathbb{R}^{H \times H}, W_2 \in \mathbb{R}^{C_t \times H}$, biases $b_1 \in \mathbb{R}^{H}, b_2 \in \mathbb{R}^{C_t}$, and the number of table types $C_t = 5$. Finally, we compute the cross-entropy loss against ground truths. To ensure an unbiased comparison, we also fine-tune TAPAS and TaBERT using the same TTC head and loss function with TUTA.

Since they didn't specify the details of dataset segmentation, we randomly split tables into train/valid/test sets under the proportion of 8/1/1. Meanwhile, having noticed previous considerations to balance samples across types, we always divide within classes and keep the type distribution in train/valid/test comparable to that of the entire dataset.

During fine-tuning, since TUTA, TAPAS, and TaBERT are all transformer-based models, we use identical hyper-parameters: with a batch size of $2$, the learning rate of $8\mathrm{e}{-6}$, and run $4$ epochs.

\end{document}